\documentclass[aps,prl,superscriptaddress,amsmath,preprint]{revtex4}
\usepackage{graphicx,multirow,longtable,amssymb}

\begin{document}


\title{Protein-mediated DNA Loop Formation and Breakdown in a Fluctuating Environment}


\author{Yih-Fan Chen*}
\affiliation{Department of Biomedical Engineering, University of Michigan, Ann Arbor, Michigan 48109}
\author{J. N. Milstein*}
\email{milsteij@umich.edu}
\affiliation{Department of Physics, University of Michigan, Ann Arbor, Michigan 48109}
\author{Jens-Christian Meiners}
\affiliation{Department of Physics, University of Michigan, Ann Arbor, Michigan 48109}
\affiliation{LSA Biophysics, University of Michigan, Ann Arbor, Michigan 48109}


\date{\today}

\begin{abstract}
Living cells provide a fluctuating, out-of-equilibrium environment in which genes must coordinate cellular function.  DNA looping, which is a common means of regulating transcription, is very much a stochastic process; the loops arise from the thermal motion of the DNA and other fluctuations of the cellular environment. We present single-molecule measurements of DNA loop formation and breakdown when an artificial fluctuating force, applied to mimic a fluctuating cellular environment, is imposed on the DNA. We show that loop formation is greatly enhanced in the presence of noise of only a fraction of $k_B T$, yet find that hypothetical regulatory schemes that employ mechanical tension in the DNA--as a sensitive switch to control transcription--can be surprisingly robust due to a fortuitous cancellation of noise effects.
\end{abstract}



\maketitle

Genes operate within a crowded cellular environment
that is constantly interacting with the encoding DNA
through various proteins that bind along its contour. We 
are only beginning to understand how chromosomal 
packing, and a crowded and constantly fluctuating cellular 
interior affects genetic regulation \cite{rivas_ellis_maniotis}; however, the 
cytoskeleton of eukaryotes already provides a startling 
example of the dynamic effects of the intracellular medium 
on cellular function \cite{cytoskeleton}. The cytoskeleton is often considered an active polymeric gel,
since its dynamic organization is driven by ATP hydrolysis, and is
known to generate mechanical stress and shear forces. These
active properties give rise to non-equilibrium effects such
as the generation of an ``effective temperature" that can
drive processes of embedded elements beyond the levels
of thermal activation. Recently, force fluctuations an order
of magnitude larger than thermal fluctuations have
been measured within the cytoskeleton \cite{gallet}. Such findings
remind us that the cellular interior is very much an active,
non-equilibrium medium, the effects of which need
to be given careful consideration in the context of gene 
expression as well. 

A regulatory mechanism that is potentially acutely sensitive to environmental noise is the fluctuation driven formation of protein-mediated DNA loops. In genomic DNA, this is a ubiquitous motif for the transcriptional control of gene expression \cite{matthews}.    The {\it lac} operon, which is responsible for efficiently metabolizing lactose in {\it E. coli} bacteria, provides a canonical example of DNA looping.  A {\it lac} repressor-mediated DNA loop is formed when tetrameric LacI protein simultaneously binds to two {\it lac} operator sites and is crucial for the repressive regulation of {\it lac} genes \cite{oehler}.  Thermal fluctuations, which generate tiny entropic forces on the order of only $k_BT/l_p\approx80$ fN,  where $l_p$ = 50 nm is the persistence length \cite{marko_blumberg}, are sufficient to form loops within the DNA, making the association rate of loops extremely sensitive to tension along the DNA molecule \cite{chen2}.

To explore the effects of environmental fluctuations on protein-mediated DNA loops, a 1316 bp dsDNA molecule with two primary {\it lac} operators spaced 305 bp apart was tethered to a coverslip and then attached to an 800 nm polystyrene microsphere.  The microsphere was then trapped within the linear region of the optical potential of a focused laser beam allowing us to apply a well defined tension to the DNA.  Details of our axial-constant force optical tweezers as well as a discussion of the DNA preparation can be found in \cite{chen1}.  Tension in the DNA was calibrated to include both the applied optical force, which is linearly proportional to the laser intensity modulated by an acousto-optic modulator (AOM), and volume exclusion effects arising from entropic interactions between the microsphere and the coverslip \cite{segall}.  The looped and unlooped states of the DNA molecule, which correspond to different axial positions of the microsphere, were measured by analyzing defocused images acquired on a CCD camera at 100 fps in the presence of 100 pM of LacI protein.  This method provides excellent temporal resolution for detecting loop formation and breakdown events with time windows as short as 300 ms.

Fluctuating forces were applied to the DNA by modulating the intensity of the trapping laser with an AOM connected to a data acquisition board and controlled by a custom LabVIEW program.  The program generated Gaussian white noise, at a sampling rate of $1/\delta t$, which was superimposed upon a set average optical force. The modulation was performed such that the force applied to the trapped microsphere was randomly chosen from a normalized Gaussian distribution of standard deviation $\sigma$.  

It is the tension along the DNA that should display white noise statistics; however,
by optically shaking the microsphere the resulting time correlation of the induced tension will effectively be low-pass filtered:
\begin{equation}\label{color}
\langle\eta(t)\eta(t')\rangle=\frac{2\alpha}{\tau_c}e^{-|t-t'|/\tau_c},
\end{equation}
where $\eta(t)$ is the DNA tension and $\alpha=\sigma^2\delta t$ is the noise strength.  The characteristic time may be approximated by $\tau_c\approx\gamma/\kappa$, where $\gamma=6\pi\eta r$ is the hydrodynamic friction coefficent of the microsphere, $\eta$ is the viscosity of the medium and r is the radius of the microsphere.  If we only consider events that happen on timescales greater than $\tau_c$ we may approximate the colored spectrum of Eq.~(\ref{color}) by purely white noise such that Eq.~(\ref{color}) reduces to
$ \lim_{\tau_c\rightarrow 0}\langle\eta(t)\eta(t')\rangle\approx 2\alpha \delta(t-t')$.  From the worm-like chain model for DNA, $\tau_c$ ranges from 5-8ms at tensions from 180-120 fN.  Therefore, if we set $\delta t$ below the cutoff imposed by $\tau_c$ (for our experiments, we fix $\delta t$ at 2 ms ($1/\delta t =500$ Hz)) then the applied fluctuations are essentially white up to frequencies $1/\tau_c$, and we may adjust the strength of the noise by tuning the width of the distribution $\sigma$.  It is more intuitive, perhaps, to consider the applied fluctuations as generating an additional energy contribution to the thermal modes of the DNA of magnitude
 $k_B T_\alpha=\sigma^2\delta t/2 \gamma$,
and to parametrize the fluctuations relative to the ambient temperature of thermal fluctuations $T_\alpha/T$.

\begin{figure}[t]
\includegraphics[scale=1.0]{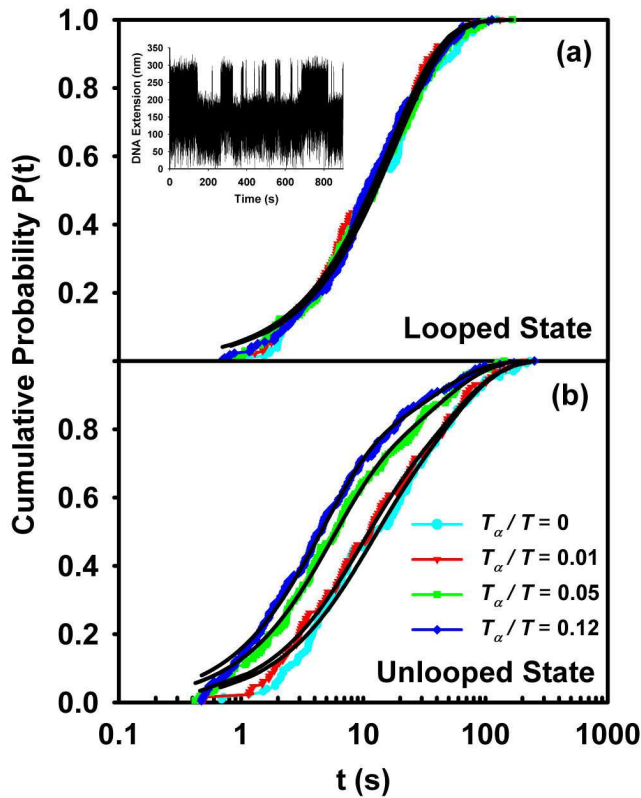}
\caption{\label{cum_fig} Experimental measurements of the (a) looped and (b) unlooped cumulative probability distributions for various noise conditions (right to left: $T_\alpha/T=0,0.01,0.05,0.12$) at a mean tension of $153$ fN.  The solid lines are exponential (biexponential) fits to the looped (unlooped) data.  The insert to (a) shows a typical raw trace of the DNA extension vs. time.}
\end{figure}

Figure~\ref{cum_fig} shows the distribution of lifetimes for the looped and unlooped states at a mean tension of 153 fN and with fluctuations of $T_\alpha/T=0,0.01,0.05,\ {\rm and}\ 0.12$.  As indicated in the figure, the lifetime of the looped state is independent of the fluctuations.  This is consistent with previous findings that the looped state is insensitive to femtonewton forces \cite{chen2}.  The lifetime of the unlooped state, however, is clearly seen to decrease as we increase the fluctuations.  At $2\sigma$, approximately 95\% of the noise distribution is accounted for.  Since volume exclusion forces are on the order of $35$ fN \cite{segall}, $\sigma=60$ fN or $T_\alpha/ T=0.12$ are the largest fluctuations we can apply to the DNA without significantly clipping the distribution.   

We fit the cumulative probability distributions using the kinetic scheme detailed in \cite{chen2} to extract loop dissociation and association lifetimes.
In summary, the looped lifetimes are simply fit by a single exponential function parametrized by the looped lifetime $\tau_L$: S(t)=$1-\exp(-t/\tau_L)$.
The unlooped kinetics, however, are more complicated, and may accurately be described by collecting all time intervals beginning with an unlooping event and ending upon the formation of a loop:
\begin{equation}\label{rate}
S_2\underset{k_1}{\overset{k_2}{\rightleftharpoons}} S_1 \overset{k_L}{\longrightarrow} L.
\end{equation}
States $S_1$ and $S_2$ arise because there are multiple unlooped sub-states available to the protein-DNA system. $S_1$ represents a state with only one occupied operator, which may either loop at rate $k_L$ to form state $L$, or remain unlooped and convert at rate $k_1$ to state $S_2$. State $S_2$ is an alternate configuration with both or neither operator occupied, which cannot form a loop, but may convert back to state $S_1$ at a rate $k_2$. The first-order kinetics results in the following biexponential function for the cumulative probability distribution:
\begin{equation}\label{loopP}
L(t)=1-\frac{1}{2\alpha}\left[c_+ e^{-t/\tau_-}-c_-e^{-t/\tau_+}\right],
\end{equation}
where $\kappa=k_2+k_1$, $\alpha=[(\kappa+k_L)^2-4k_2k_L]^{1/2}$, $c_\pm=(\kappa-k_L\pm\alpha)$ and the time constants are defined as  $\tau_\pm=2/(\kappa+k_L\pm\alpha)$. From this fit equation we are able to extract the unlooped lifetime $\tau_u=1/k_L$.  In accord with the constant force results \cite{chen2}, the interconversion rates $k_2$ and $k_1$ were  found to be essentially independent of the applied tension.
\begin{figure}[t]
\includegraphics[scale=1.0]{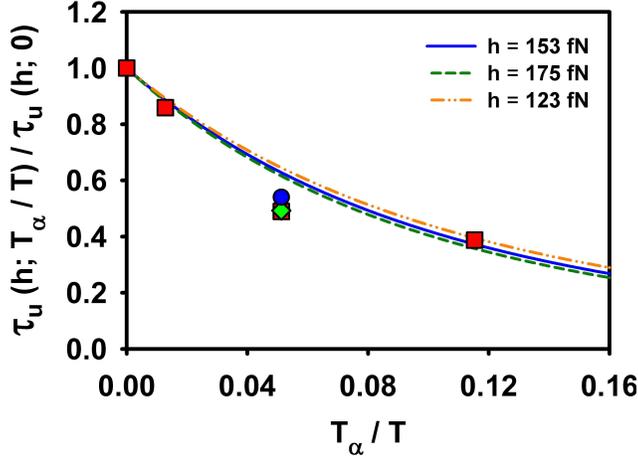}
\caption{\label{tau_vs_sigma} Normalized unlooped lifetimes as a function of applied noise.  The (square) data points were taken at a mean tension of $h=153$ fN.  The (diamond) and (circle) data points were taken at $h=123$ and $175$ fN respectively.  Theoretical curves are shown for $ h=123,153,175$ fN.}
\end{figure}

Figure \ref{tau_vs_sigma} displays the unlooped lifetimes as a function of the noise $T_\alpha/T$ normalized to the zero noise lifetime, $\tau_u(h;T_\alpha/T)/\tau_u(h;0)$, about a mean applied tension of $153$ fN.  The results demonstrate that fluctuations do indeed drive loop formation and coalesce nicely with previous observations that femtonewton forces can radically affect the rate at which LacI-mediated DNA loops form \cite{chen2}.  Our results imply that an access contribution from environmental fluctuations of only 5\% of the ambient thermal temperature $T$ can double the rate at which DNA loops form, which is only a fraction of that which the cytoskeleton can induce upon an embedded polymer \cite{gallet}.  This stochastic mechanism might, therefore, provide an alternate `noisy' means for mechanical control of genetic transcription. 


Although the rate at which DNA loops form is quite sensitive to environmental fluctuations, our data also show that this sensitivity is practically independent of the mean applied tension in the DNA.  A separate measurement of the loop formation rate as a function of mean tension, $h=123,153,$ and $175$ fN, collected at a constant applied noise, $T_\alpha/T=0.05$, reveals a striking observation: the normalized lifetimes, $\tau_u(h;0.05)/\tau_u(h;0)$, are constant with an approximate value of $0.5$ irrespective of the average tension $h$ in the molecule (see Fig. \ref{tau_vs_sigma}).  As we will show, this could allow regulatory schemes that are based upon protein-mediated DNA loops to display a significant level of robustness to noise.  

\begin{figure}[b]
\includegraphics[scale=.75]{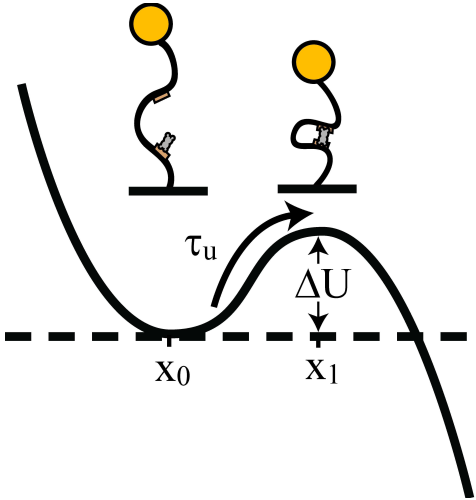}
\caption{\label{barrier_fig} The loop association process can be modeled by diffusion over a barrier.  The unlooped lifetime $\tau_u$ is given by the average time it takes for the DNA to diffuse from the equilibrium position $x_0$, within the energy landscape, to the top of the energy barrier, of magnitude $\Delta U$ at $x_1$, where it can escape to infinity and form a loop. }
\end{figure}

One might imagine that the increase in looping occurs because the fluctuations allow the random sampling of a low tension regime.  However, any increase in looping gained by reducing the tension is negated by the equally likely sampling of high tension fluctuations, requiring us to consider the role of fluctuations in more detail.  We begin with an effective Langevin equation for the motion of the tethered particle system:
\begin{equation}\label{diffeq}
\gamma \frac{dx}{dt} =-\frac{dU(x;h)}{dx}+\xi(t).
\end{equation}
The reaction variable $x(t)$ describes diffusion along the energy landscape provided by $U(x;h)$.  The stochastic term $\xi(t)$ accounts for thermal fluctuations and is modeled as a white noise source with zero mean, i.e.,
$\langle\xi(t)\xi(t')\rangle=2\lambda\delta(t-t')$,
with $\lambda=\gamma k_B T$.

We choose the following phenomenological form for the energy barrier: 
\begin{equation}\label{pot}
U(x;h)=\frac{1}{2}ax^2-\frac{1}{3}b x^3+h x.
\end{equation}
The harmonic term contains the cost of bending the DNA, while the cubic term, although somewhat arbitrary, is the simplest contribution that can give rise to an energy barrier between a second equilibrium state, which is here assumed to be the looped configuration, see Fig. \ref{barrier_fig}.  It should be noted that the parameter $a$  is quite different from the spring constant $\kappa$ of a worm-like chain model since $a$ must parameterize the full three dimensional search of the polymer to form a loop within this lower dimensional, effective energy landscape.  The linear term represents the force $h$ that we apply with optical tweezers to stretch the DNA, which effectively modulates the energy barrier by tilting the energy landscape and, therefore, increasing or decreasing the barrier height $\Delta U$.  Note that we are not attempting to account for the unlooping process, since the looped state, as our data reveal, is not sensitive to the forces we apply.

\begin{figure}[t]
\includegraphics[scale=1.0]{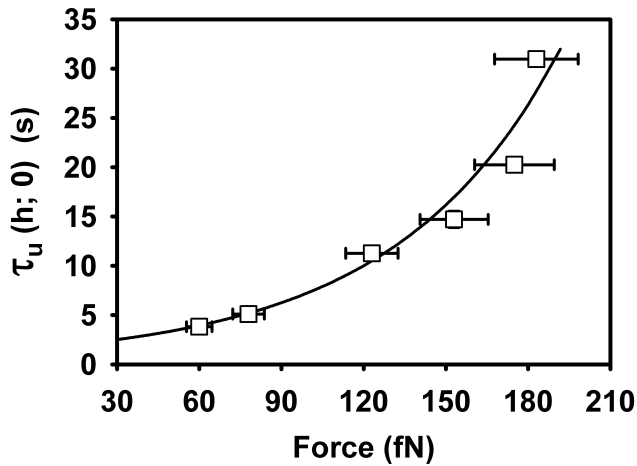}
\caption{\label{lifetime_fig} Unlooped lifetime $\tau_u$ as a function of a constant applied force.  The solid line is the theoretical fit, given by the Kramers relation of Eq.~(\ref{kfit}) to the data points (squares).  The coefficients of the potential $U(x;h)$ are $a=4.64\times 10^{-5}\ {\rm N/m}$ and $b=739\ {\rm N/m^2}$.  }
\end{figure}

An exact formula for the mean passage time \cite{gardiner} across the energy barrier from $x_0$ to $x_1$ is given by
\begin{equation}\label{exact_fpt}
\tau_u=\frac{\gamma}{k_B T}\int_{x_0}^{x_1}\hspace{-2mm}dx \exp\left(\frac{U(x;h)}{k_B T}\right)\int_{-\infty}^x \hspace{-2mm}dy\exp\left(\frac{-U(y;h)}{k_B T}\right).
\end{equation}
If the potential barrier $\Delta U$ is large compared to $k_B T$, then Eq.~(\ref{exact_fpt}) can be expanded about the vicinity of $x_0$ and $x_1$ to yield the Kramers formula for the unlooped lifetime
\begin{equation}
\tau_u=\frac{2\pi\gamma}{\sqrt{U''(x_0;h)|U''(x_1;h)|}}\exp \left[\frac{\Delta U}{k_B T}\right],
\end{equation}
where $\Delta U=U(x_1;h)-U(x_0;h)$.  From Eq.~(\ref{pot}), we can rewrite this relation as
\begin{equation}\label{kfit}
\tau_u=\frac{2\pi\gamma}{\sqrt{a^2+4bh}}\exp \left[\frac{(a^2+4bh)^{3/2}}{6 b^2 k_B T}\right].
\end{equation}
We must now determine the coefficients $a$ and $b$ that parameterize our model potential (Eq.~(\ref{pot})). We do this by an iterative least-squares fit of the lifetimes, given by Eq.~(\ref{exact_fpt}), to our constant force (zero fluctuation) data, the result of which is shown in Fig.~\ref{lifetime_fig}.  We have found the difference in the resulting fit parameters to be negligible between the exact, Eq.~(\ref{exact_fpt}), and approximate Kramers relation, Eq.~(\ref{kfit}).  Moreover, the zero tension ($h=0$) energy barrier  $\Delta U=a^3/6b^2=7.4 k_B T$, from this fit, is in remarkable agreement with the energy cost for loop formation predicted by 
an elastic rod model of DNA, $\Delta U= 7.7 k_B T$ \cite{wilson}.

Since we add noise to the system by linearly modulating the tension applied to the DNA, we may incorporate this additional noise by modifying the correlations of the stochastic source $\xi(t)$ such that $\lambda=\gamma k_B T_E$, where we have introduced the effective temperature
\begin{equation}
T_E=T(1+T_\alpha/T).
\end{equation}
With the replacement $T\rightarrow T_E$ in Eq.~(\ref{kfit}) we are able to account for the effects of noise on the unlooped lifetimes $\tau_u(h;T_\alpha/T)$ at a mean tension $h$.  This model gives excellent agreement with our experimental measurements of the looping lifetime $\tau_u(h;T_\alpha/T)$ as a function of noise (see Fig.~\ref{tau_vs_sigma}).  Furthermore, it predicts a robustness to noise similar to what we observe experimentally (see Fig.~\ref{tau_vs_sigma}).  The theory predicts a fairly constant ratio $\tau_u(h;0.05)/\tau_u(h;0)=0.65,0.63,$ and $0.62$ at a mean tension  $h=123,153,$ and $175$ fN, respectively.

Our results suggest how a force dependent genetic switch that employs DNA looping to regulate transcription could operate stably within a noisy environment.  For instance, consider a regulatory element controlled by the formation of a DNA loop at a basal rate $k_{t_1}$ under constant tension $h_1$.  A regulatory signal could be provided by a change in tension, $h_2$, such that the loop formation rate is now a factor of $p$ times the basal rate, i.e. $k_{t_2}/k_{t_1}=p$.  If we assume that the tension felt by the DNA fluctuates around the targets $t_1$ and $t_2$ such that we have two new looping rates $\tilde{k}_{t_1}$ and $\tilde{k}_{t_2}$, our results imply that $\tilde{k}_{t_1}/k_{t_1}=\tilde{k}_{t_2}/k_{t_2}$, which means that $\tilde{k}_{t_2}/\tilde{k}_{t_1}=p$, so that the expression signal is unaffected by the noise.

We have demonstrated how noise from thermal and environmental fluctuations drives protein-mediated DNA loop formation, yet leaves the loops unaffected once formed. Environmental fluctuations only a fraction of the size of thermal fluctuations in the DNA can greatly enhance the rate at which these loops form. We interpret these results with a fluctuating barrier model that can quantitatively explain and predict our measurements. This model is based on the previously demonstrated sensitivity of loop formation to static mechanical tension, which led to the suggestion that cells may utilize tension to regulate transcription through mechanical pathways, as opposed to the more commonly considered biochemical ones \cite{chen2}. Based on our new observations, we may now postulate the feasibility of an alternate mechanical regulatory mechanism that uses environmental fluctuations as a means to control transcription. The rate enhancement of fluctuations might, in fact, explain why a simple loop results in several hundred fold repression {\it in vivo} 
\cite{oehler}, even though the looped and unlooped states have roughly equal
lifetimes {\it in vitro}.  Furthermore, we have shown that the sensitivity of loop formation to fluctuations is insensitive to baseline static mechanical tension, and have demonstrated how this feature can lead to a robustness in regulatory function. 


\acknowledgements{*Y. F. Chen and J. N. Milstein contributed equally to this work.
We thank Jason Kahn for the LacI protein and David Wilson for help with the modeling.  This work was supported by NIH, Grant No. GM65934, and NSF FOCUS, Grant No. 0114336.}

\end{document}